\documentclass[aps,prb,twocolumn,amsmath,amssymb,superscriptaddress]{revtex4-1}
\usepackage{graphicx}
\usepackage{dcolumn} 
\usepackage{graphics}
\usepackage{bbm}
\usepackage{amsmath,amsfonts,amssymb}
\usepackage{wrapfig}
\usepackage[pdftex]{color}

\begin{document}
\title{Cryogenic Control Architecture for Large-Scale Quantum Computing}

\author{J. M. Hornibrook}
\affiliation{ARC Centre of Excellence for Engineered Quantum Systems, School of Physics, The University of Sydney, Sydney, NSW 2006, Australia.} 
\author{J. I. Colless}
\affiliation{ARC Centre of Excellence for Engineered Quantum Systems, School of Physics, The University of Sydney, Sydney, NSW 2006, Australia.} 

\author{I. D. Conway Lamb}
\affiliation{ARC Centre of Excellence for Engineered Quantum Systems, School of Physics, The University of Sydney, Sydney, NSW 2006, Australia.} 

\author{S. J. Pauka}
\affiliation{ARC Centre of Excellence for Engineered Quantum Systems, School of Physics, The University of Sydney, Sydney, NSW 2006, Australia.} 

\author{H. Lu}
\affiliation{Materials Department, University of California, Santa Barbara, California 93106, USA.}
\author{A. C. Gossard}
\affiliation{Materials Department, University of California, Santa Barbara, California 93106, USA.}

\author{J. D. Watson}
\affiliation{Department of Physics, Purdue University, West Lafayette, Indiana 47907, USA.}
\affiliation{Birck Nanotechnology Center, Purdue University, West Lafayette, Indiana 47907, USA.}

\author{G. C. Gardner}
\affiliation{School of Materials Engineering, Purdue University, West Lafayette, Indiana 47907, USA.}
\affiliation{Birck Nanotechnology Center, Purdue University, West Lafayette, Indiana 47907, USA.}

\author{S. Fallahi}
\affiliation{School of Materials Engineering, Purdue University, West Lafayette, Indiana 47907, USA.}
\affiliation{Birck Nanotechnology Center, Purdue University, West Lafayette, Indiana 47907, USA.}

\author{M. J. Manfra}
\affiliation{Department of Physics, Purdue University, West Lafayette, Indiana 47907, USA.}
\affiliation{Birck Nanotechnology Center, Purdue University, West Lafayette, Indiana 47907, USA.}
\affiliation{School of Materials Engineering, Purdue University, West Lafayette, Indiana 47907, USA.}
\affiliation{School of Electrical and Computer Engineering, Purdue University, West Lafayette, Indiana 47907, USA.}

\author{D. J. Reilly$^\dagger$}
\affiliation{ARC Centre of Excellence for Engineered Quantum Systems, School of Physics, The University of Sydney, Sydney, NSW 2006, Australia.} 

\begin{abstract}
Solid-state qubits have recently advanced to the level that enables them, in principle, to be scaled-up into fault-tolerant quantum computers. As these physical qubits continue to advance, meeting the challenge of realising a quantum machine will also require the engineering of new classical hardware and control architectures with complexity far beyond the systems used in today's few-qubit experiments. Here, we report a micro-architecture for controlling and reading out qubits during the execution of a quantum algorithm such as an error correcting code. We demonstrate the basic principles of this architecture in a configuration that distributes components of the control system across different temperature stages of a dilution refrigerator, as determined by the available cooling power. The combined setup includes a cryogenic field-programmable gate array (FPGA) controlling a switching matrix at 20 millikelvin which, in turn, manipulates a semiconductor qubit. 
\end{abstract}

\maketitle

Realising the classical control system of a quantum computer is a formidable scientific and engineering challenge in its own right\cite{Van_Meter,Devitt}. The hardware that comprises the control interface must be fast relative to the timescales of qubit decoherence, low-noise so as not to further disturb the fragile operation of qubits, and scalable with respect to physical resources, ensuring that the footprint for routing signal lines or the operating power does not grow rapidly as the number of qubits increases\cite{Fowler,DiV}. As solid-state quantum processors will likely operate below 1 kelvin\cite{Reed_Nature,3D_transmon,Loss:1998via,Majorana_Scheme}, components of the control system will also need to function in a cryogenic environment, adding further constraints. 

Similar challenges have long been addressed in the satellite and space exploration community\cite{nasa}, where the need for high-frequency electronic systems operating reliably in extreme environments has driven the development of new circuits and devices \cite{Extreme_electronics}. Quantum computing systems, on the other hand, have to date largely relied on brute-force approaches, controlling a few qubits directly via room temperature electronics that is hardwired to the quantum device at cryogenic temperatures. 

\begin{figure}
\centering
\includegraphics[scale=0.49]{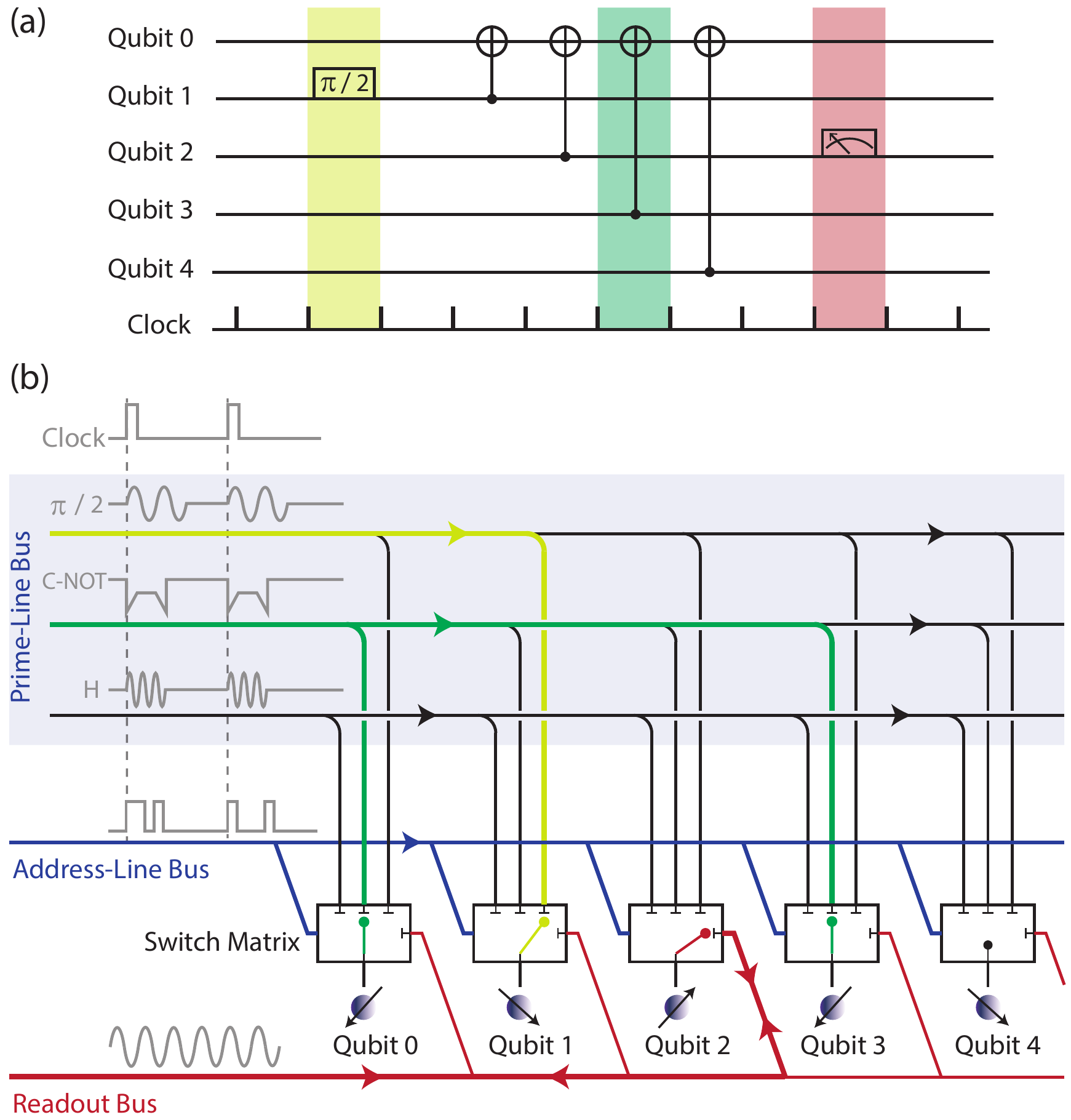}
\caption{\label{fig:PLAL}{\bf `Prime-line / Address-line' (PL/AL) architecture that separates prime analog waveforms, used to manipulate qubits, from the addressing data used to select qubits. } {\bf (a)} An example quantum algorithm shown using quantum circuit notation. The highlighted clock cycles include a single-qubit rotation (yellow), a two-qubit gate (green) and readout operation (red). {\bf(b)} Prime-lines corresponding to a universal gate set are routed to qubits via a switching matrix controlled by the address-lines. Coloured paths correspond to the highlighted clock cycles in (a).}
\end{figure}

Here we present a control architecture for operating a cryogenic quantum processor autonomously and demonstrate its basic building blocks using a semiconductor qubit. This architecture addresses many aspects related to scalability of the control interface by embedding multiplexing sub-systems at cryogenic temperatures and separating the high-bandwidth analog control waveforms from the digital addressing needed to select qubits for manipulation. Our demonstration comprises a commercial field-programmable gate array (FPGA) operating at 4 kelvin and controlling a microwave signal switching matrix at 20 mK, which then interfaces with a quantum dot device. Bringing these sub-systems together in the context of our control architecture suggests a path for scale-up of control hardware needed to manipulate the large numbers of qubits in a useful quantum machine.

\section{Control Micro-architecture}
Our control micro-architecture takes advantage of the universality of quantum gates, which allows for arbitrary logic operations to be realised using a small set of repeated single- and two-qubit unitaries applied in sequence. At the level of physical qubits in the solid-state, whether they are spins \cite{Yacoby2qubit}, transmons \cite{3D_transmon}, or quasi-particles \cite{majorana}, these elemental gate operations amount to applying calibrated electrical waveforms to a particular qubit or pair of qubits each clock cycle as determined by a quantum algorithm. 

A key aspect of our control architecture is the separation of these analog `prime waveforms', which are typically pulses at microwave frequencies, from the digital qubit addressing information that determines which waveform is directed to which qubit, at a particular point in the code. In comparison to brute-force approaches, this scheme lifts the need of having a separate waveform generator and transmission line for each qubit, taking advantage of a small universal gate set that uses the same analog waveforms over-and-over throughout the algorithm. As realistic qubits will inevitably include variations in their physical parameters, the control architecture must also incorporate means of calibrating and adjusting the response of the qubit to the control waveforms, as described below.

Our `prime-line / address-line' (PL/AL) architecture is shown schematically in Fig. \ref{fig:PLAL}, where we have drawn part of a circuit for implementing a quantum error correcting surface code \cite{Bravyi,Raussendorf}. Precisely timed analog prime waveforms, generated at cryogenic or room temperature, propagate cyclicly on a high-bandwidth prime-line bus that is terminated with a matched impedance at a location in the system where heat can be dissipated.  The quantum algorithm is then executed exclusively via the digital address-line bus, selecting qubits and qubit pairs to receive the appropriate prime waveform at the correct clock cycle in the circuit. Readout proceeds in a similar way, with the digital address bus selecting a particular qubit (or readout device) for interfacing with analog readout circuitry such as a chain of amplifiers and data converters. 

\section{Implementation of the Control Architecture}
\begin{figure}
\centering
\includegraphics[scale=0.46]{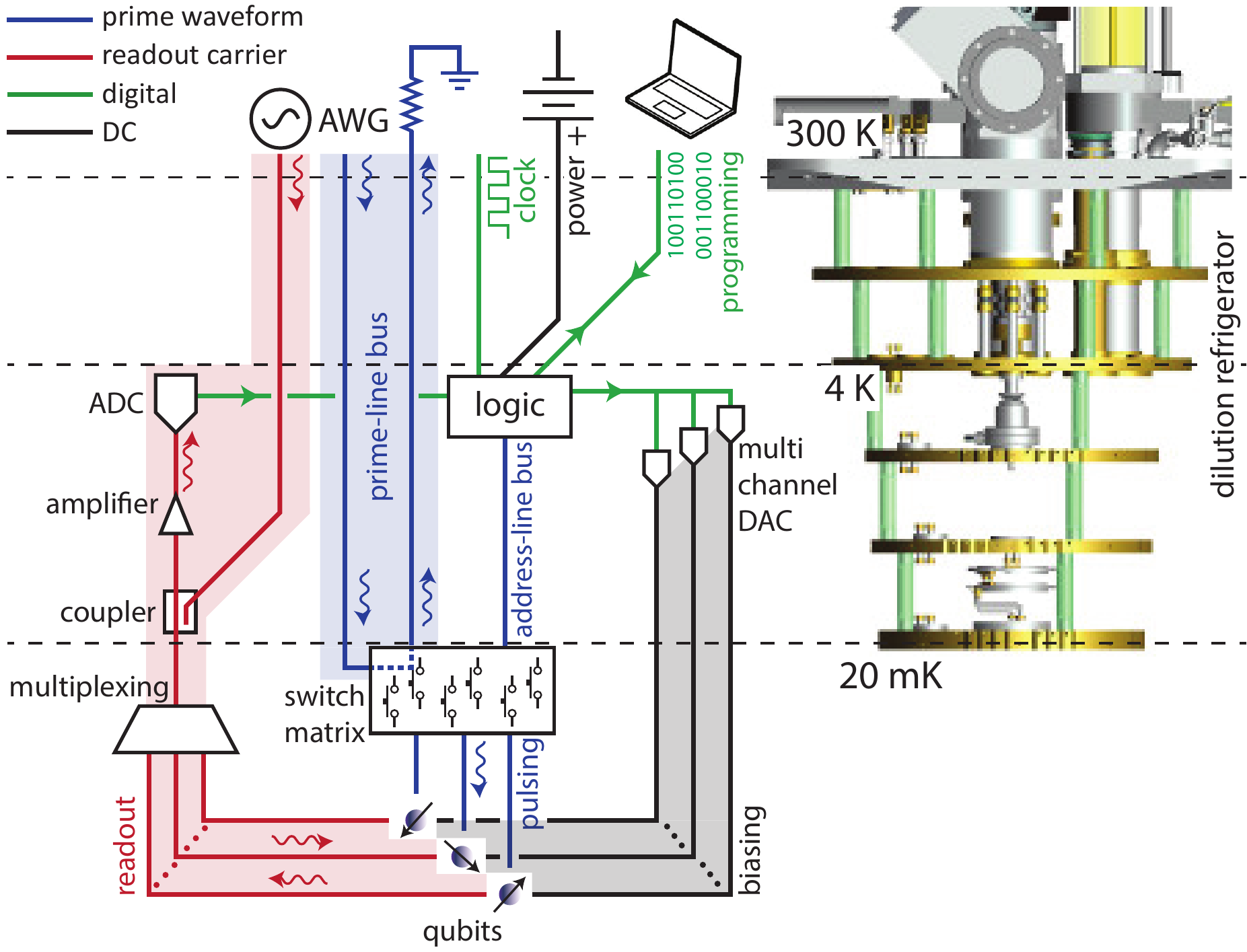}
\caption{\label{fig:DRstages}{\bf Schematic of a control micro-architecture that distributes sub-systems across the various temperature stages of a dilution refrigerator, depending on the available cooling power} (image is of a Leiden Cryogenics CF450). A millikelvin switching matrix, on the same chip as the qubit device or close to it, steers a small number of control pulses to qubits using addressing information from cryogenic logic at 4 K. The cryogenic logic also interfaces with multiplexed readout and digital-to-analog converters. The 4 K stage typically has a cooling power $\sim$ 1 W, with the 20 mK stage having less than 1 mW.}
\end{figure}

Realising our PL/AL architecture requires integrating multi-component control and readout hardware with the quantum system of qubits fabricated on a chip. Owing to the large number of qubits that are likely to be needed for quantum computation and the timescales involved in their control, there are advantages to locating sub-systems of the control architecture at cryogenic temperatures, either on-chip with the physical qubits, or in close proximity and connected via integrated multi-chip modules \cite{MCM_ribbonSC} and compact transmission lines. Aspects of the control system will however, generate significant heat or fail to function at the millikelvin temperatures needed for qubit operation. The competing constraints of interconnect density, heat generation, signal latency, footprint, and noise performance suggest a control architecture that is distributed across a cryostat, taking advantage of the significantly different thermal budgets available at each temperature stage. This distributed architecture is illustrated in Fig. \ref{fig:DRstages}, where control sub-systems are positioned at different temperature stages of a cryogen-free dilution refrigerator. Below we describe and provide a basic demonstration of these sub-systems. 

\section{Switching Matrix}
The key sub-system underpinning the control micro-architecture is a switching matrix, or routing technology that steers the prime waveforms to particular qubits based on a digital address. This technology is ideally located in close proximity to the qubits to avoid latency and synchronization challenges that arise when signals propagate over length-scales comparable to the electromagnetic wavelength (typically centimetres for quantum control waveforms). Physically integrating the switching matrix and qubit system has the further advantage of significantly reducing the wiring and interconnect density by making use of lithography (or multi-chip module packaging) to provide connection fan-out. In this way we envisage a switching matrix that receives multiplexed data on a small number of transmission lines and decodes this address data to operate large numbers of parallel switches (see Fig. 2). Multiplexing of this kind will likely be essential for operation in cryogenic environments where large numbers of parallel transmission lines add a sizeable heat load when carrying signals between stages that are at different temperatures. The use of superconducting materials is key as these can dramatically reduce the cross-section and thereby thermal load of transmission lines without degrading electrical performance \cite{MCM_ribbonSC}.

A switching matrix with elements that act as variable impedances can also be configured to enable the amplitude and phase of the prime waveforms to be individually adjusted before arriving at each qubit. By incorporating a calibration routine or feedback scheme, this approach can be used to account for the variation in physical parameters that will inevitably occur with systems comprising large numbers of qubits.
 
Various technologies appear suitable for constructing such a switching matrix, including semiconducting devices \cite{Eriksson,Smith,Smith2}, mechanical systems \cite{pzt,pzt2}, and superconducting logic \cite{RQL}. For qubit technologies built from semiconductors \cite{Colless_PRL,Majorana_Scheme}, field-effect based devices are ideally suited owing to their sub-nanosecond switching-speed, gigahertz transmission bandwidth, low dissipation, small footprint, cryogenic compatibility, and opportunity for integration with qubits. Below we demonstrate the operation of such devices using GaAs high electron mobility transistor (HEMT) circuits, configured as a switching matrix with variable amplitude and phase response.

\subsection{HEMT Switching Elements}
\begin{figure}
\centering
\includegraphics[scale=0.47]{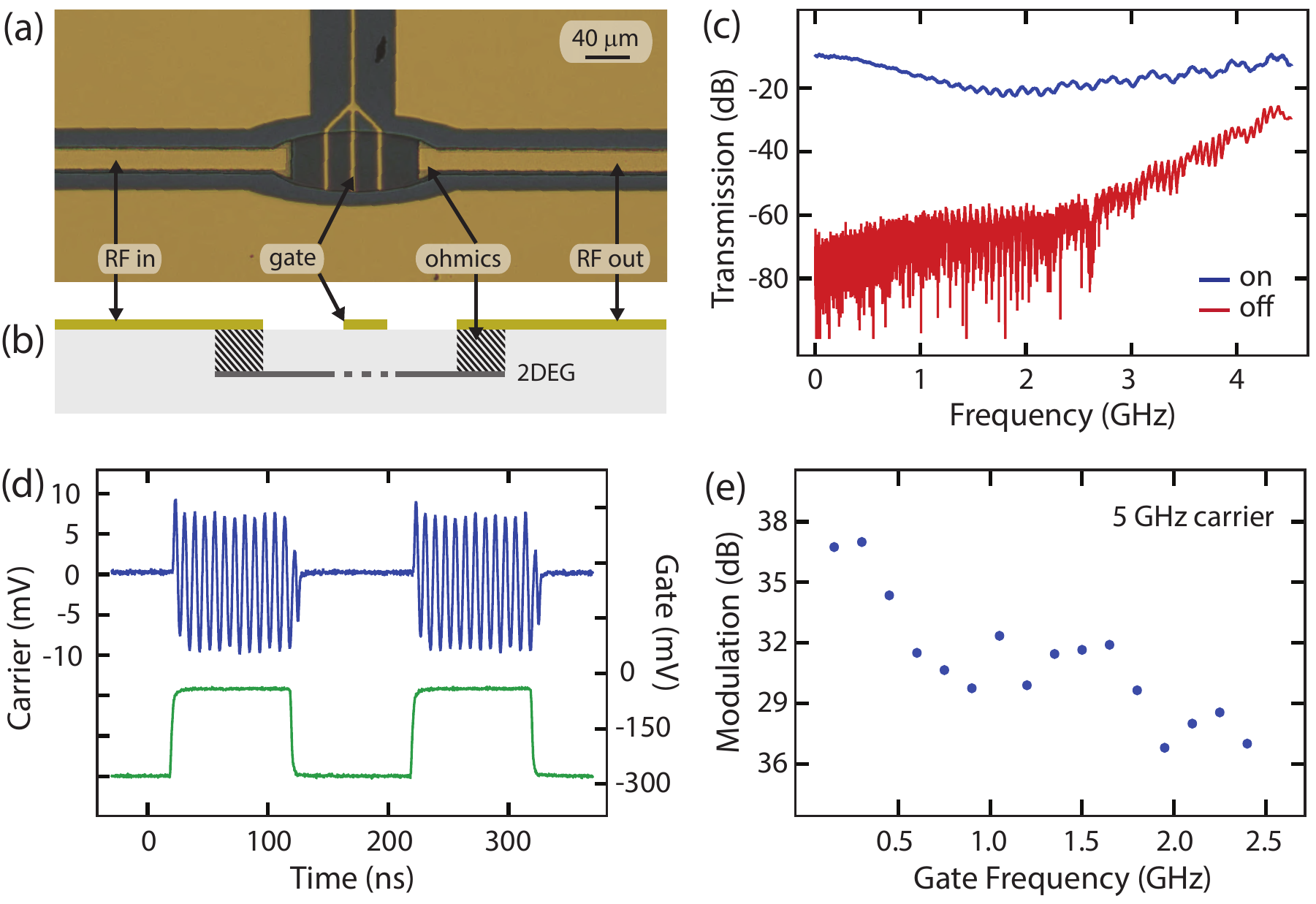}
\caption{\label{fig:HEMTswitch} {\bf Characterisation of a HEMT switch as a building block for the PL/AL architecture.} {\bf(a)} Microscope photograph of the device fabricated on GaAs/Al$_{0.3}$Ga$_{0.7}$As heterostructure. {\bf(b)} Schematic cross-section showing the coplanar line diverted through the 2DEG. A negative voltage (-300 mV) on the top gate increases the impedance of the switch, reflecting the input signal. {\bf(c)} Transmission as a function of frequency for the on (blue) and off (red) state. {\bf(d)} Example of time-domain response. When the gate voltage (green) is zero, the 120 MHz sine wave provided at the switch input is propagated to the output (blue), and not otherwise. {\bf (e)} Modulating a carrier signal through the 2DEG with a sinusoidal gate voltage creates sidebands. The amplitude of the sidebands as a function of frequency indicates a 1 - 2 ns switching time.}
\end{figure}

A prototype HEMT-style microwave switch based on a GaAs/AlGaAs heterostructure is shown in Fig. \ref{fig:HEMTswitch}(a,b). In the on-state, the switch is configured to have a characteristic impedance of $\sim$ 50 $\Omega$, owing to its coplanar waveguide (CPW) geometry. Prime waveforms are fed to and from the HEMT two-dimensional electron gas (2DEG) via eutectic ohmic contacts and TiAu planar transmission lines. In the off-state a negative voltage applied to the TiAu top gate pinches-off the electron gas channel, reflecting the prime waveform signal due to the large impedance of the HEMT relative to the characteristic impedance of the $\sim$ 50 $\Omega$ feedline. The transmission response of the switch is shown in Fig. \ref{fig:HEMTswitch}(c), with an on/off ratio (OOR) above 40 dB in the frequency range 0 - 2.5 GHz, suitable for control of spin qubits \cite{Reilly:2010gp}. For these prototype devices a large insertion loss of 10-20 dB is observed, owing mostly to the resistance of the ohmic contacts, which deviates from 50 $\Omega$. Precise control of the contact resistance and capacitance using ion-implantation can overcome this limitation and also dramatically shrink the footprint of these devices \cite{ohmics1, ohmics2}.

The time-domain response of the switch is demonstrated by amplitude modulating an applied 120 MHz constant wave tone, as shown in Fig. \ref{fig:HEMTswitch}(d). To determine the maximum switching time of the HEMT we modulate a 5 GHz carrier tone with a sinusoidal waveform applied to the gate and measure the depth of modulation as a function of gate frequency, as indicated in Fig. \ref{fig:HEMTswitch}(e). For these prototype devices the switching time is of order 1 ns.

\subsection{Capacitive Switching Elements}
\begin{figure}
\centering
\includegraphics[scale=0.43]{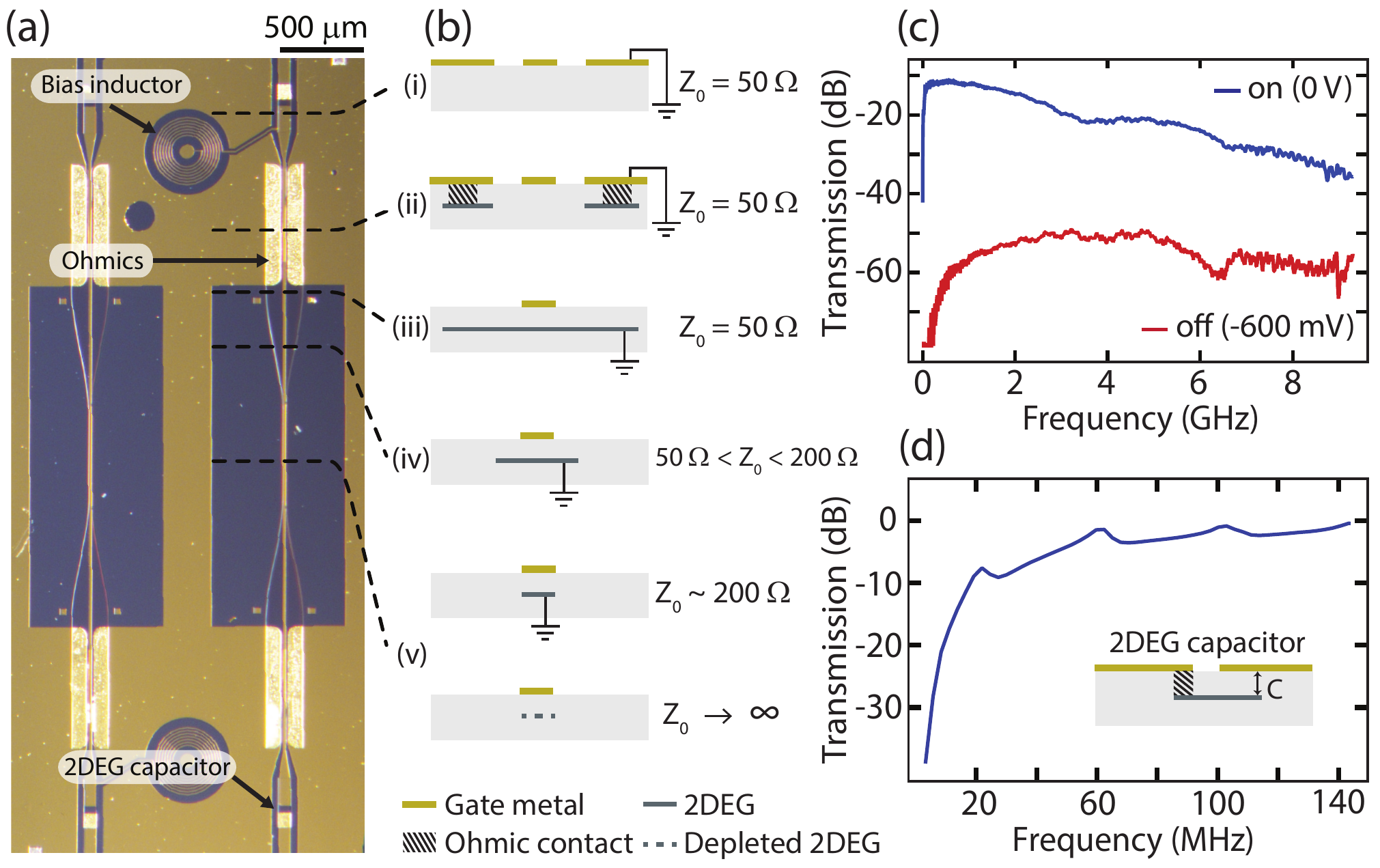}
\caption{\label{fig:CAPswitch}{\bf A switch design that produces an impedance mismatch by depleting the transmission line ground plane.} Shown is an image {\bf(a)} and cross-sections {\bf(b)} of the device. The input is a coplanar line (i) which transitions to a microstrip using the 2DEG as the ground plane (ii,iii). The ground plane is tapered so that it lies beneath the signal track (iv,v). When a negative voltage is applied to the signal track, the ground plane is depleted (v) and the impedance mismatch reflects the input signal. {\bf(c)} Transmission through the switch in the on (blue) and off (red) states. {\bf(d)} Frequency response of capacitors formed using surface gates and 2DEG as a parallel plate (inset).}
\end{figure}

Microwave switching devices based on the depletion of an electron gas also enable a new capacitive mode of operation. In this configuration the CPW feedline transitions to a microstrip geometry by contacting the electron gas to the planar ground planes using ohmic contacts, as illustrated in Fig. \ref{fig:CAPswitch}(a,b). The two conductors in the microstrip transmission line are thus constructed using the top gate and electron gas as ground. This device can act as a reflective switch by depleting the effective ground plane using a negative bias on the gate. Depletion reduces the capacitance between the conductors of the microstrip and modulates the device impedance. 
Transmission through the switch is shown in Fig. \ref{fig:CAPswitch}(c) in the on (blue) and off (red) state, with an OOR greater than 25 dB for 0 - 8 GHz. To the best of our knowledge, a switching device based on a depleted ground plane has not been reported previously.

The switch is capacitively coupled to the input and output ports, with a planar spiral inductor at one port forming a bias tee to provide the dc gate voltage needed to deplete the electron gas. In place of a planar interdigitated capacitor, we make use of the GaAs heterostructure to provide a low footprint parallel plate capacitor, formed between the CPW central track and the electron gas, as shown in the inset to Fig. \ref{fig:CAPswitch}(d). The frequency response of this capacitor is shown in Fig. \ref{fig:CAPswitch}(d). 

The capacitance-based switch has improved performance at higher frequency than the HEMT-based switch, although it has a larger footprint due to both the length of line needed for adiabatic tapering from 50 to 200 $\Omega$ and for the coplanar-to-microstrip transition. This improvement stems from the absence of a gate structure, which in the HEMT switch capacitively couples the source and drain contacts, even in the off state. The required footprint is reduced significantly in an all-microstrip circuit that is designed to operate at a characteristic impedance close to 200 $\Omega$.

\subsection{2:2 Switch Matrix}
\begin{figure}
\centering
\includegraphics[scale=0.56]{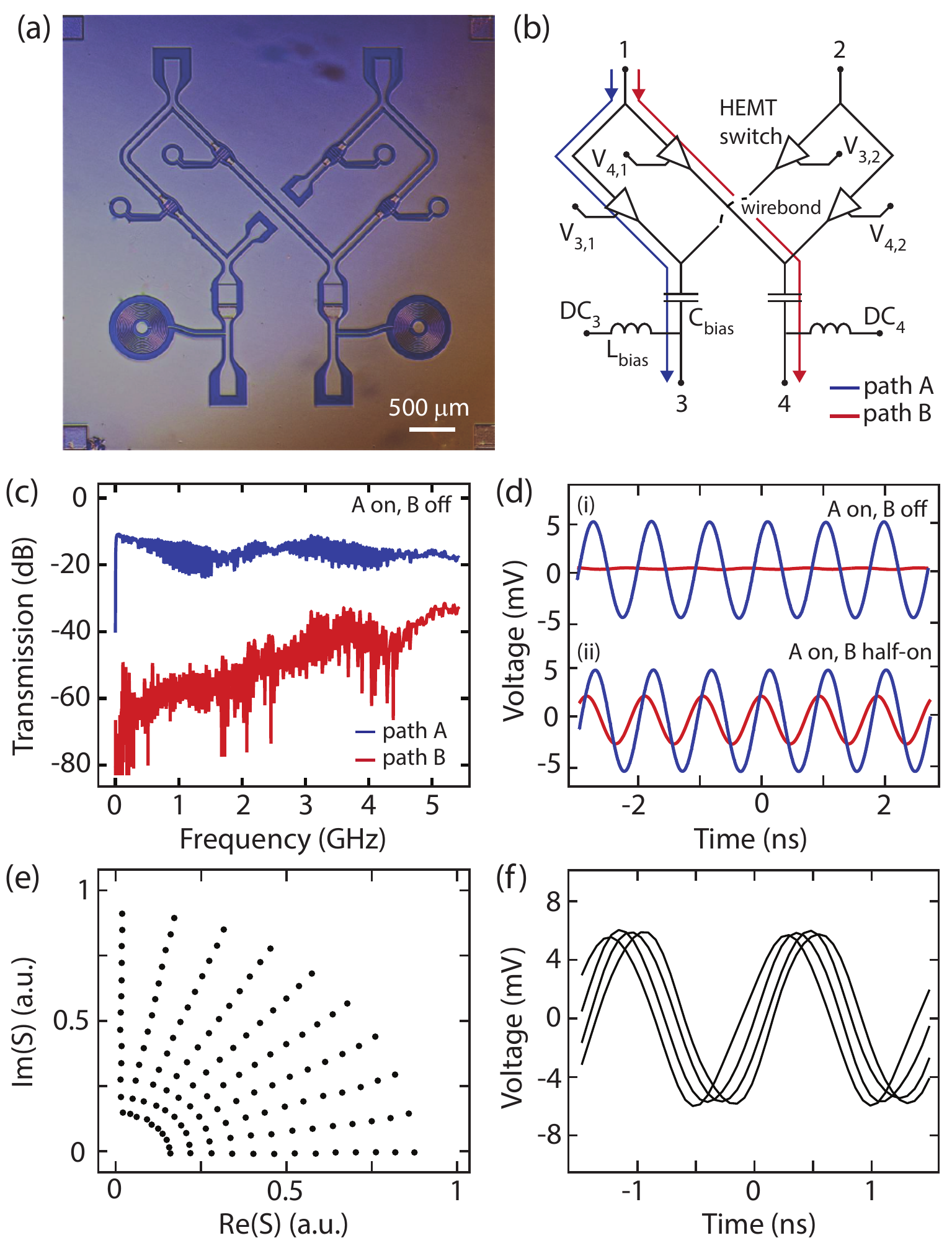}
\caption{\label{fig:diplexer}{\bf Small scale 2-input, 2-output switching matrix based on HEMT switches, with on-chip bias tees for quantum dot operation.} Device image is shown in {\bf(a)} with associated circuit diagram in {\bf(b)}. {\bf(c)} Transmission measurement with path A (blue) in the on-state and path B (red) in the off-state. {\bf(d)} Voltage output with a 1 GHz input tone where path A is in the on-state and path B is (i) off, and (ii) half-on. {\bf(e)} An example of IQ modulation, implemented by feeding the input ports of the 2:2 matrix with signals that have a 90$^{\circ}$ phase offset. Arbitrary amplitude and phase is produced at the output (data shown in figure) by selecting the appropriate $V_{i,j}$ (see main text). {\bf(f)} Example voltage output for one of the constant amplitude quarter circles in (e).}
\end{figure}

We demonstrate cryogenic operation of a prototype routing matrix based on HEMT switches with two input and two output ports. A magnified image of the device is shown in Fig. \ref{fig:diplexer}(a) with associated schematic in (b). Each input port is split and connected to each output port via a switch so that the transmission parameters $S_{ij}$ of the device are controlled by the respective gate voltages $V_{i,j}$. The output ports include bias tees, which are needed for use with qubits based on semiconductor quantum dots.  Bias tees are constructed using planar spiral inductors and 2DEG-based capacitors as illustrated in the inset of Fig. \ref{fig:CAPswitch}(d).

Operation of this switching matrix is demonstrated by comparing the transmission of signals as a function of frequency for path A (blue) and path B (red), as indicated in Fig. \ref{fig:diplexer}(b). The response through both paths when path A is on ($V_{3,1}$=0) and path B is off ($V_{4,1}$ = -500 mV) is shown in Fig. \ref{fig:diplexer}(c). The corresponding time-domain response for a 1 GHz tone is shown in Fig. \ref{fig:diplexer}(d)(i). We observe a negligible ($<$ 0.05 dB) change in the response of one path when the other is path is switched from the on state to the off state. An advantage of semiconductor-based switching elements is their ability to be configured as variable impedances,  producing arbitrary amplitude output, as shown in Fig. \ref{fig:diplexer}(d)(ii). 

We also demonstrate basic IQ modulation using our switching matrix by applying rf tones at both inputs with a 90$^{\circ}$ phase offset between them. The 90$^{\circ}$ shift can be produced by a length of transmission line (with narrowband response) or as a separate quadrature prime waveform. The output waveform at angular frequency $\omega$ is $A \sin \omega t + B \cos \omega t = R \sin (\omega t + \phi)$, where the magnitude $R$ and phase $\phi$ are determined by the amplitudes $A$ and $B$, controlled by the gate voltages $V_{i,j}$. After the calibration function $R, \phi = \mathcal{F} (V_{i,j})$ is generated once, we can select the appropriate $V_{i,j}$ to produce a tone with arbitrary phase and amplitude in the first quadrant of the complex plane, as shown in Fig. \ref{fig:diplexer}(e). The corresponding voltage output along a quarter circle of constant amplitude is shown in Fig. \ref{fig:diplexer}(f). By controlling the amplitude and phase shift using the integrated switching matrix, the connection between each qubit and the prime line bus can be specifically adjusted to compensate for the inevitable variation in parameters between physical qubits \cite{footnotecalibration_qubit_swtich_together}.

\section{Cryogenic Logic}
For controlling and programming the switching matrix via the address bus, we envisage a layer of fast, classical logic that serves as an interface between the physical qubits and compiled quantum algorithm (comprising mostly an error correcting code). This layer of classical logic is also needed for executing various automatic sequences associated with fast feedback for qubit stabilisation, readout signal conditioning, or open-loop error suppression\cite{Viola,Yacobyfeedback}. For controlling a large-scale quantum computer there are many advantages to locating this classical logic and associated data converters close to the qubits, inside a dilution refrigerator. In comparison to room temperature based control systems, cryogenic operation results in an enhanced clock speed, improved noise performance, reduced signal latency, and larger bandwidth. Some of these aspects stem from the ability to make use of compact superconducting transmission lines and interconnects at cryogenic temperatures.

The choice of technology for constructing this layer of classical control is largely dictated by the qubit coherence times, control signal bandwidth, and the number of simultaneous qubits under control. With a convergence of solid-state qubit coherence times now approaching 1 millisecond \cite{Pla,Bluhm,Reagor}, present day CMOS-based FPGAs or application specific integrated circuits (ASICs) operating at 4 kelvin are a viable control platform. Higher performance control systems that are likely to be realised in the longer term include technologies based on InP devices \cite{InP}, SiGe BiCMOS \cite{Cressler,You}, and superconducting flux logic \cite{RQL}.

For the basic demonstration of the PL/AL scheme considered here the classical logic is implemented using a commercial FPGA (Xilinx Spartan-3A) that we have made operational at the 4 K stage of a dilution refrigerator. To achieve cryogenic operation the FPGA chip was mounted on a custom, cryogenic printed circuit board that includes components which vary little in their parameters at cryogenic temperatures \cite{CollessRSI, CollessRSI2}. Power and clock signals to the FPGA are adjusted for cryogenic operation using room temperature sources and a semi-rigid coax line is configured for sending serial commands, with reprogramming of the low temperature array occurring via a dedicated ribbon cable. With the FPGA mounted at the 4 K stage we measure an idle power dissipation of $\sim$ 30 mW, with negligible increase during dynamic logic operations for the simple code executed here. We estimate a dynamic power dissipation of $\sim$ 100 mW for computational operations that use most of the gates in the Spartan-3 array (further details of cryogenic operation of FPGAs are given elsewhere \cite{FPGA_paper}). The FPGA is programmed to interpret serial communication and output a 3.3 V signal on selected pins to activate prime waveform routing in the switching matrix. These outputs are combined with a negative voltage provided from room temperature via a cold resistive adder so that the switching matrix gates receive -50 mV for the on-state and -380 mV for the off-state voltage.

\section{Semiconductor Qubit Control}
We combine the building-blocks of our micro-architecture described above, to demonstrate that a semiconductor qubit can feasibly be controlled autonomously without introducing additional noise or heating to the quantum system. The qubit is a GaAs double quantum dot configured as a charge or spin qubit in the few-electron regime. These qubits are commonly controlled using dc-pulse waveforms on the gates to rapidly manipulate the energy levels of the quantum dots \cite{Petta_science05}. A typical setup connects a waveform generator to each gate using a separate high bandwidth coaxial cable and bias tee. 

\begin{figure}
\includegraphics[scale=0.46]{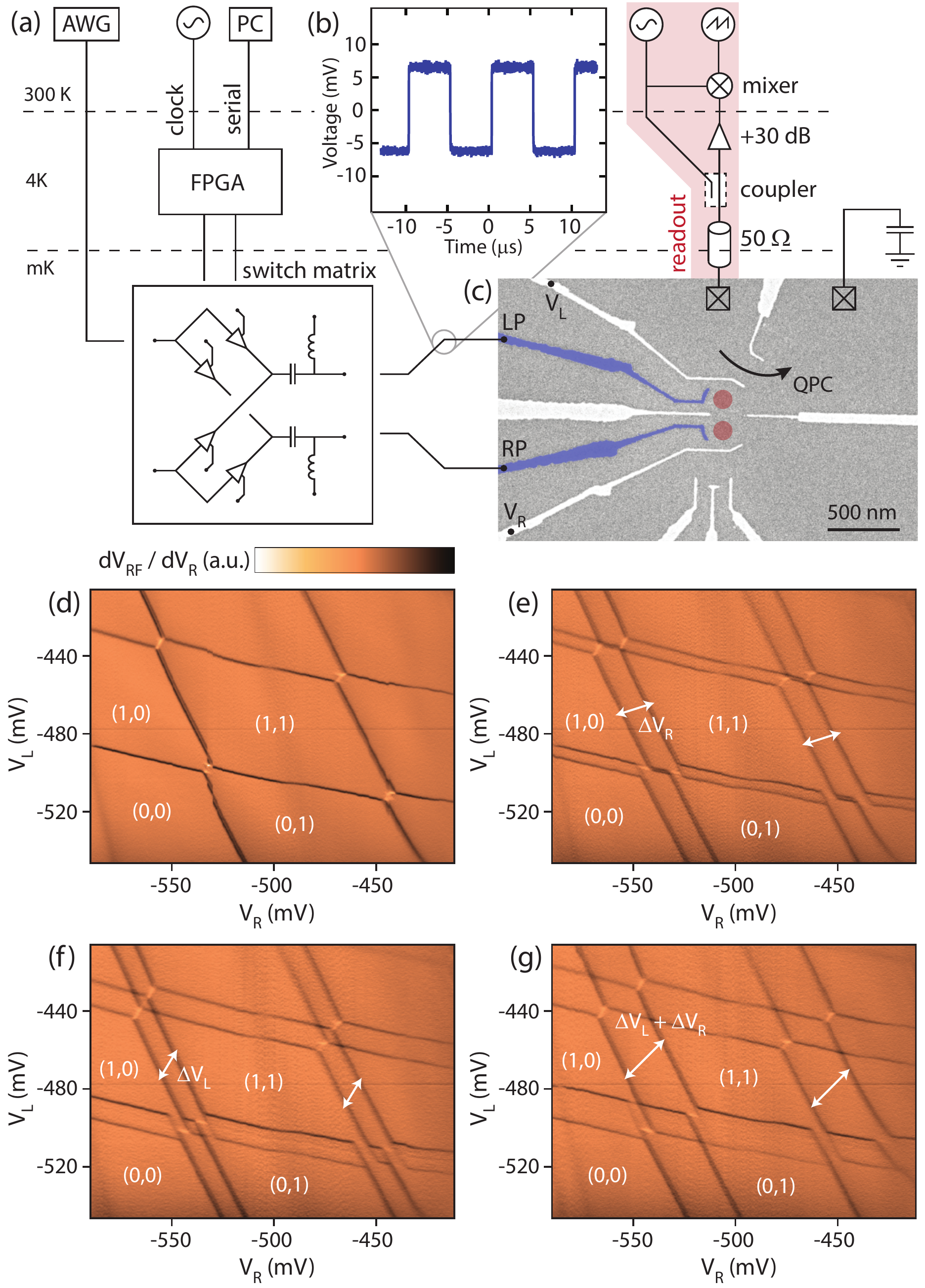}
\caption{\label{fig:combined}{\bf Simple implementation of the micro-architecture introduced in Fig. \ref{fig:DRstages}}. {\bf(a)} Experimental setup for measuring a double quantum dot, using a cryogenic FPGA to steer pulses via a millikelvin switching matrix. Charge-state readout is performed using an rf-QPC. {\bf(b)} Switching matrix output showing a 100 kHz square wave directed to plunger gates of the quantum dot. {\bf(c)} Micrograph of the quantum dot device. The shaded gates, labelled LP and RP, are connected to the switching matrix output. {\bf(d-g)} Charge sensing of the double quantum dot in the few-electron regime, with electron occupancy indicated by the labels (m, n). The colour axis is the derivative of the sensing signal $V_{rf}$ with respect to $V_R$. When the FPGA-controlled switching matrix blocks waveforms {\bf(d)}, a standard double dot stability diagram is detected. When the square wave is directed to either LP {\bf (e)}, RP {\bf(f)} or both {\bf(g)}, copies of the stability diagram appear (see text). These measurements demonstrate that the double dot potential can be controlled autonomously by the switching matrix and cold FPGA.}
\end{figure}

For this demonstration we connect a single coaxial cable from a waveform generator at room temperature to the input of the 2:2 switching matrix, with the two matrix output ports connected to the two plunger gates $LP$ and $RP$ of the double dot, as shown schematically in Fig. \ref{fig:combined}(a).  The waveform generator produces a prime waveform consisting of a 100 kHz square wave (shown in Fig. \ref{fig:combined}(b)) which is then steered by the 4 kelvin FPGA by opening and closing switches in the matrix depending on commands sent from room temperature. 

The charge state of the double dot is sensed using an rf quantum point contact \cite{Reilly:2007ig,Hornibrook_APL}, which provides a readout signal $V_{rf}$ as a function of the gate voltages $V_L$ and $V_R$ indicated in (c). With both switches of the matrix set to the off state, a standard charge stability diagram is detected indicating that the off state provides sufficiently high isolation between input and output ports, as shown in Fig. \ref{fig:combined}(d) \cite{low_freq_jitter}. 

Sending a command to the cold FPGA allows the prime waveform to be directed to the left, right, or both plunger gates, producing two copies of the charge stability diagram. These copies appear because, on the timescale of the readout, a square wave with 50\% duty-cycle configures the double dot in two distinct charge states that are offset from one another by the voltage $\Delta V_R$ or $\Delta V_L$, as shown in Fig. \ref{fig:combined}(e-g) \cite{Petta_PRB05}. In comparison to data taken on the bare quantum dot, we are unable to detect any additional noise or an increase in the electron temperature (which is of order 100 mK) when configuring the charge-state using the cryogenic FPGA and switching matrix.

\section{Discussion}
Our simple demonstration of a multi-component control architecture provides a path for scaling up the classical support system needed for operating a large-scale quantum computer. Aspects of this demonstration will also likely find immediate use in improving the performance of few-qubit experiments using electron spins in quantum dots. For example, in using the switching matrix to produce multiple out-of-phase copies of control waveforms, crosstalk can be suppressed by cancelling the voltage that is capacitively coupled to neighbouring surface gates \cite{Blanvillain}. Using the switching matrix as a high frequency cryogenic multiplexer will also enable the automated testing and characterisation of many devices in a given cool-down experiment. In the longer term, our micro-architecture can be extending to allow additional functionality of the switching matrix, providing qubit control frequency correction by using the HEMTs as mixers, or as cryogenic adder circuits that reduce the complexity or resolution needed for biasing surface gates that define quantum dots. 

At the layer of classical logic, our demonstration shows that commercial FPGA devices can be configured to work at cryogenic temperatures and are compatible with controlling qubits in close proximity. Beyond the control architecture presented here, the use of cold, low-latency classical logic will likely improve the performance of feedback systems generally needed for adaptive measurement, quantum state distillation, and error correction protocols.  Given the power dissipation inherent to such control systems however, improvements in cryogenic refrigeration technology, similar to what has been achieved in rare-event physics \cite{CUORE}, will likely be needed to enable large-scale quantum information processing.

\section{Conclusion}
We have proposed a micro-architecture for the control of a large-scale quantum processor at cryogenic temperatures. The separation of analog control prime waveforms from the digital addressing needed to select qubits offers a means of scaling this approach to the numbers of qubits needed for a computation. To demonstrate the feasibility of our scheme we have shown that a semiconductor qubit can be controlled using a cryogenic FPGA system and custom switching matrix for steering analog waveforms at low temperature. We anticipate that integrated, autonomous control systems of this kind will be increasingly important in the development and demonstration of fault tolerant quantum machines.

\section{Methods and Materials}
The fabrication of GaAs switching elements follows similar procedure to quantum dot devices (allowing easy integration). The mesa is wet etched using sulphuric acid, before Au/Ge/Ni ohmic contacts are thermally evaporated and annealed at 470 degrees for 100 seconds. The final metal layer is thermally evaporated TiAu (10 nm / 100 nm). The device (Fig. \ref{fig:combined}(c)) is an electrostatically defined double quantum dot on GaAs/Al$_{0.3}$Ga$_{0.7}$As heterostructure (carrier density 2.4 $\times$ 10$^{-15}$ m$^{-2}$, mobility 44 m$^2$ / Vs at 20 K). 

\section{Acknowledgements}
We thank B. Smith, D. Tuckerman, D. Wecker, K. Svore, C. M. Marcus, L. DiCarlo, L. P. Kouwenhoven, and M. Freedman for useful conversations. Devices were fabricated at ANFF-NSW. This research was supported by the Office of the Director of National Intelligence, Intelligence Advanced Research Projects Activity (IARPA), through the Army Research Office grant W911NF-12-1-0354, the Australian Research Council Centre of Excellence Scheme (Grant No. EQuS CE110001013), and Microsoft Research.

\section{Author contributions}
D.J.R. devised the micro-architecture, J.M.H. fabricated the switches and switching matrix, M.J.M., J.D.W., G.G., and S.F. grew the heterostructure for the switches. J.M.H. and S.J.P. performed the experiment on the switches, J.M.H., J.I.C., I.D.CL., and S.J.P. performed the experiment on the quantum dot with FPGA control. H.L. and A.C.G. grew the heterostructure for the quantum dot. J.M.H. and D.J.R. wrote the manuscript.

$\dagger$ Corresponding author, email: david.reilly@sydney.edu.au 

\bibliographystyle{naturemag}


\begin{thebibliography}{10}
\bibitem{footnotecalibration_qubit_swtich_together}
\bibinfo{title}{{\textnormal Calibration of the switch response and qubit can
  be performed at the same time, measuring the qubit evolution as a function of
  switch gate voltage}} .
\bibitem{low_freq_jitter}
\bibinfo{title}{{\textnormal A very small amount of jitter in the charge
  transitions can be seen due to coupling of the rf-QPC carrier to the gates}}
  .
\end{thebibliography}

\end{document}